\documentclass[runningheads]{llncs}
\usepackage[T1]{fontenc}
\usepackage{graphicx}
\usepackage{xcolor}
\usepackage{hyperref}
\usepackage{booktabs}
\usepackage{multirow}
\usepackage{wrapfig}
\usepackage[algo2e, ruled]{algorithm2e}
\usepackage{todonotes}
\usepackage{float}
\usepackage{graphicx}
\usepackage{amsmath, amssymb, amsfonts}

\newcommand{\eg}{\textit{e.g.}}
\newcommand{\ie}{\textit{i.e.}}

\newcommand{\R}{\mathbb{R}}
\newcommand{\E}{\mathbb{E}}

\begin{document}
\title{Anatomically-Controllable Medical Image Generation with Segmentation-Guided Diffusion Models}
\titlerunning{Anatomically-Controllable Segmentation-Guided Diffusion Models}
%
\author{Nicholas Konz\inst{1}\textsuperscript{*} \and Yuwen Chen\inst{1} \and Haoyu Dong\inst{1} \and Maciej A. Mazurowski\inst{1,2,3,4}}
\authorrunning{N. Konz et al.}
\institute{Department of Electrical and Computer Engineering, Duke University, NC, USA\and
Department of Radiology, Duke University, NC, USA \and
Department of Computer Science, Duke University, NC, USA \and
Department of Biostatistics \& Bioinformatics, Duke University, NC, USA  \\ *\textit{corresponding author} \email{nicholas.konz@duke.edu}
}

%
\maketitle              

\begin{abstract}
Diffusion models have enabled remarkably high-quality medical image generation, yet it is challenging to enforce anatomical constraints in generated images. 
To this end, we propose a diffusion model-based method that supports anatomically-controllable medical image generation, by following a multi-class anatomical segmentation mask at each sampling step. We additionally introduce a \textit{random mask ablation} training algorithm 
to enable conditioning on a selected combination of anatomical constraints while allowing flexibility in other anatomical areas.
We compare our method (``\textbf{SegGuidedDiff}'') to existing methods on breast MRI and abdominal/neck-to-pelvis CT datasets with a wide range of anatomical objects. Results show that our method reaches a new state-of-the-art in the faithfulness of generated images to input anatomical masks on both datasets, and is on par for general anatomical realism. 
Finally, our model also enjoys the extra benefit of being able to adjust the anatomical similarity of generated images to real images of choice through interpolation in its latent space. SegGuidedDiff has many applications, including cross-modality translation, and the generation of paired or counterfactual data. Our code is available at \url{https://github.com/mazurowski-lab/segmentation-guided-diffusion}.

\end{abstract}

\section*{Introduction}

\begin{figure}[htbp]
   \centering
   \includegraphics[width=0.6\linewidth]{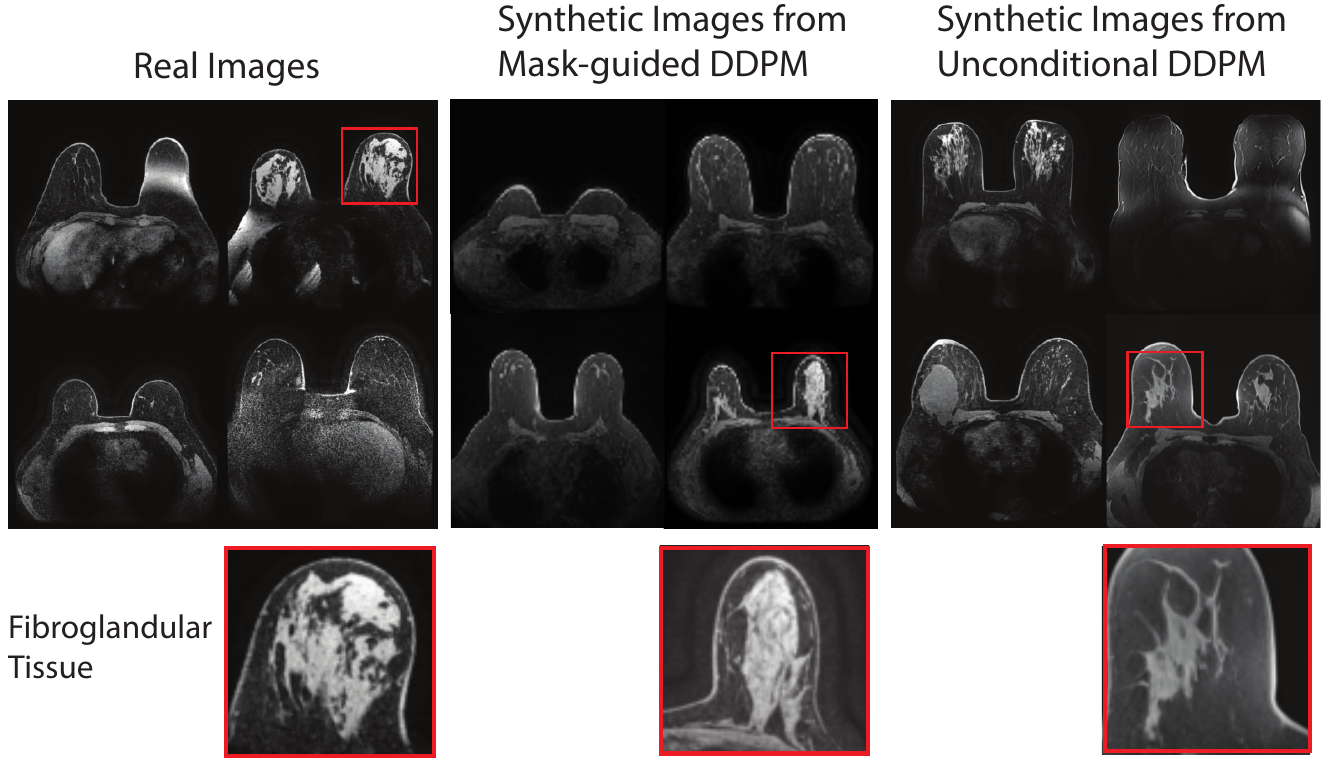}
   \caption{Standard diffusion models (right) can fail to create realistic tissue even if the overall image appears high-quality, motivating our segmentation-guided model (center).
   }
   \label{fig:eg_imgs}
\end{figure}

Denoising diffusion probabilistic models \cite{ho2020denoising} (DDPMs, or just ``diffusion models'') have shown extensive applications in medical image analysis \cite{kazerouni2023diffusion} due to their ability to generate high-quality, high-resolution images, such as for direct image generation \cite{pinaya2022brain,khader2023denoising}, image segmentation \cite{wolleb2022diffusionseg}, anomaly detection \cite{pinaya2022fast,wolleb2022diffusion}, cross-modality image translation \cite{lyu2022conversion}, and image denoising \cite{gong2023pet}.
However, standard generative models like DDPMs can still fail to create anatomically plausible tissue (Fig \ref{fig:eg_imgs}), and such anatomy is not precisely customizable. Our proposed solution is to incorporate anatomical information as a prior for image generation via a segmentation mask for different types of tissue, organs, etc., providing the network with a more direct learning signal for anatomical realism.

Generating an image from a mask (a.k.a. \textit{semantic synthesis}) is a type of image-to-image translation task. Existing translation works include GAN-based \cite{yang2019unsupervised,choi2018stargan,cao2023deep} and diffusion-based models \cite{wolleb2022diffusion,stablediffusion,controlnet}, yet these do not directly enforce precise pixel-wise anatomical constraints.
Recent works introduced fine-tuning large pre-trained text-to-image latent diffusion models (LDMs) for segmentation-conditioning on natural images \cite{controlnet,wang2022piti}, but we found that these adapt poorly to medical images (Sec. \ref{sec:expcompare}). Instead, we propose the first diffusion model for generating medical images from anatomical segmentations, which is assisted by it being an image-space diffusion model trained from scratch.
Image-space-based diffusion models are especially amenable for strict segmentation guidance because a conditioning mask can be used as-is for every small step of the denoising process, with no conversion to some abstract latent space as in certain LDMs \cite{controlnet,wang2022piti,stablediffusion} (or GANs) where precise spatial guidance may be lost.

Segmentation-guided generation would be even more flexible if only certain object classes could be constrained in an input mask, while others are free to be inferred by the model. This opens up further applications such as generating images from incomplete masks (\cite{tian2023partial}),
the generation of anatomically paired/registered data, ``counterfactual'' analysis of existing annotated data, and others.
To solve this, we propose a \textbf{mask-ablated training} strategy to provide the model with all possible combinations of missing classes in masks during training, teaching it to make such inferences when generating new images. Notably, we also demonstrate how interpolating within the latent space of our mask ablated-trained model enables generating images with adjustable anatomical similarity to some real image (Sec. \ref{sec:expmaskabla}).

In summary, we introduce a diffusion model, ``\textbf{SegGuidedDiff}'' that can flexibly and precisely generate medical images according to anatomical masks. We evaluate our model's ability to generate images conditioned on a range of anatomical objects of interest for breast MRI and neck-to-pelvis CT, where it outperforms state-of-the-art mask-conditional generative models in its faithfulness to input anatomical masks, and is on par for general anatomical realism (Sec. \ref{sec:expcompare}).
Our code is publicly released at \url{https://github.com/mazurowski-lab/segmentation-guided-diffusion} with a focus on usability on any dataset, along with a dataset of paired ``pre-registered'' generated breast MRIs.

\section{Method}

\subsection{A Brief Review of Diffusion Models}
Denoising diffusion probabilistic models \cite{ho2020denoising} (DDPMs, or diffusion models for short) are a type of generative latent variable model that learns to sample from some data distribution $p(x_0)$ ($x_0\in\R^n$) by defining a stochastic process that gradually converts the data to noise---the \textit{forward} process $q(x_t|x_{t-1})$---and learning to reverse this process via a learned denoising process $p_\theta (x_{t-1}|x_t)$, where $\theta$ is the model parameters. Data is generated by iteratively sampling from $p_\theta (x_{t-1}|x_t)$, beginning with a Gaussian noise sample $x_T\sim p(x_T)$, for $t=T-1,\ldots, 0$ (we use $T=1000$) until an image $x_0$ is recovered. 

Any forward process step can be written explicitly as $x_t=\sqrt{\overline{\alpha}_t}x_0 + \sqrt{1- \overline{\alpha}_t}\epsilon$ where $\epsilon\sim\mathcal{N}(0, I_n)$, and $\alpha_t:= 1- \beta_t$ given the variance of the additive pre-scheduled noise $\beta_t$, and $\overline{\alpha}_t:=\prod_{s=1}^t\alpha_s$. DDPMs can be trained by the usual evidence lower bound (ELBO) maximization, which can be approximately optimized in a relatively simple form by training a network $\epsilon_\theta(x_t,t)$ to predict the noise $\epsilon$ added to each datapoint $x_0$ for various time steps $t$, with the loss
    $L=\E_{x_0,t,\epsilon} \left [||\epsilon - \epsilon_\theta(x_t,t)||^2\right]= \E_{x_0,t,\epsilon} \left [||\epsilon - \epsilon_\theta(\sqrt{\overline{\alpha}_t}x_0 + \sqrt{1- \overline{\alpha}_t}\epsilon,t)||^2\right]$,
which has proven to be the typically superior DDPM loss in practice \cite{nichol2021improved}.

\subsection{Adding Segmentation Guidance to Diffusion Models}
\label{sec:segguidance}

Rather than sampling from the unconditional distribution $p(x_0)$, our goal is to condition the generation of some $c$-channel image $x_0\in\R^{c\times h\times w}$ to follow some multi-class anatomical mask $m\in\{0,\ldots, C-1\}^{h\times w}$, where $C$ is the number of classes (including background), or in other words, sample from $p(x_0|m)$. While modifying the data likelihood $p(x_0|m)$ to be mask-conditional does not alter the noising process $q(x_t|x_{t-1})$, it does modify the reverse process $p_\theta(x_{t-1}|x_t,m)$ and noise-predicting network $\epsilon_\theta$. Propagating this to the ELBO results in a loss of
\begin{equation}
    \label{eq:lossmask}
    L_m=\E_{(x_0,m),t,\epsilon} \left [||\epsilon - \epsilon_\theta(\sqrt{\overline{\alpha}_t}x_0 + \sqrt{1- \overline{\alpha}_t}\epsilon,t|m)||^2\right]
\end{equation}
for training our model, where each training image $x_0$ has some paired mask $m$. We propose to implement this simply by concatenating $m$ channel-wise to the network input at every denoising step, \ie, modifying the network to have an additional input channel as $\epsilon_\theta(x_t,t|m):\R^{(c+1)\times h\times w}\rightarrow \R^{c\times h\times w}$, which can be any image-to-image model (see Sec. \ref{sec:implement}). We use the DDIM algorithm \cite{song2021ddim} for fast, yet high-quality sampling.

This simple method generates images that are very faithful to input masks (Fig. \ref{fig:samples}, Table \ref{tab:faithful}), because the denoising process is conditioned on the mask at each of its many gradual steps, allowing the network to follow the masks because they provide helpful spatial information that is directly correlated with the optimal denoised model output that minimizes the loss.

\subsection{Mask-Ablated Training and Sampling}
\label{sec:mask_abla}

Given that our model is mask-guided, the quality of these masks is important; a generated image may be misleading if the input mask is not fully annotated, known as the \textit{partial label problem }\cite{tian2023partial} in medical image analysis. This is because the model may assume that un-annotated objects should not be present in the output image whatsoever (associating the missing/zero pixel labels as background or some other object), when in reality we may desire for the model to simply ``fill in''/infer the unprovided objects. 

To alleviate this problem, we propose a \textbf{mask-ablated training} (MAT) strategy (Algorithm  \ref{alg:training}), which provides examples of masks with various numbers and combinations of classes removed for the model to learn to generate images from during training. This can be thought of as a form of self-supervised learning of anatomical object representations (somewhat analogous to MAE \cite{he2022masked}).
We set all $2^{C-1}$ of these possible combinations of classes being removed from a given mask in training to occur with equal probability so that the model can handle each equally, although any other balancing of these probabilities for different object classes could be used. Finally, we note that our MAT algorithm is immediately applicable to \textit{any} mask-conditional generative model.

\SetKwRepeat{Do}{do}{while}
\begin{algorithm2e}[H]
  \caption{Segmentation-guided model training with mask ablation.} \label{alg:training}
  \KwIn{number of mask classes $C$, dataset $p(x_0,m)$.}
    \Repeat{converged}{
      $x_0,m\sim p(x_0,m)$\\
      \For{$c = 1,\ldots, C-1$}{
        $\delta \sim \mathrm{Bernoulli}(0.5)$\\
        \uIf{$\delta = 1$}{
            $m[m=c] = 0$\\
        }
      }
      $\epsilon\sim\mathcal{N}(0,I_n)$; $t \sim \mathrm{Uniform}(\{1, \dotsc, T\})$\\
      $x_t = \sqrt{\bar\alpha_t} x_0 + \sqrt{1-\bar\alpha_t}\epsilon$\\
      Update $\theta$ with $\nabla_\theta \left\| \epsilon - \epsilon_\theta(x_t, t|m) \right\|^2$\\
    }
\end{algorithm2e}

\section{Datasets}
\label{sec:data}
\textbf{Breast MRI:}
Our first dataset is a 100-patient subset of the Duke Breast Cancer MRI dataset 
\cite{saha2018machine}. We use all 2D image slices from the fat-saturated gradient echo T1-weighted pre-contrast sequence, with a train/test split of 70/15 patients, resulting in $\sim$12000/2500 slice images per split. We also keep a held-out training set of 15 patients for additional experiments. All images have full segmentation annotations for (1) breast, (2) blood vessels (BV), and (3) fibroglandular/dense tissue (FGT) provided at \cite{Lew2024APA}. Notably, the FGT and BV have very high variability in shape, size, and other morphological characteristics, posing a challenge for generative models to realistically capture.
\textbf{CT Organ:}
Our second dataset is a 40-patient subset of neck-to-pelvis and abdominal CT scans from \cite{rister2020ct}, with segmentation annotations for liver, bladder, lungs, kidney, and bone. This results in a train/test split of $\sim$11000/2100 2D slice images, given a patient-wise split of 24/8, as well as a held-out training set of 8 patients. All generative models are trained on the training sets, and the auxiliary segmentation network, introduced next, is trained on the held-out training sets.

\section{Experiments}
\subsubsection{Training, Architecture and Implementational Details.}
\label{sec:implement}
All images are resized to $256\times 256$ and normalized to $[0, 255]$. We use a UNet architecture \cite{ronneberger2015u} for the denoising model $\epsilon_\theta$, modified to take two channels (image and mask) as input; see Appendix A for additional training and architecture details.

\subsection{Comparison to Existing Image Generation Models}
\label{sec:expcompare}

We next compare our segmentation-guided diffusion model (``\textbf{SegGuidedDiff}'' for short) to state-of-the-art segmentation-conditional image generation models. These are SPADE \cite{park2019SPADE}, a GAN-based model that uses spatially-adaptive normalization layers, and ControlNet \cite{controlnet}, a recent method for adding spatial conditioning to large pretrained text-to-image diffusion models. Training/implementation details for each are in Appendix A.1; note that we use the standard implementation of SPADE rather than the recent brain MRI SPADE model \cite{fernandez2022brainspade,fernandez_3d_2024} because its modifications to SPADE do not apply to our setting (see Appendix A.1), and that we follow ControlNet's guidelines for adapting it to medical images.
We show example generated images from all models in Fig. \ref{fig:samples} (using masks randomly sampled from the test set); more are provided in Appendix B.

\begin{figure}[htbp]
   \centering
    \includegraphics[width=0.44\linewidth]{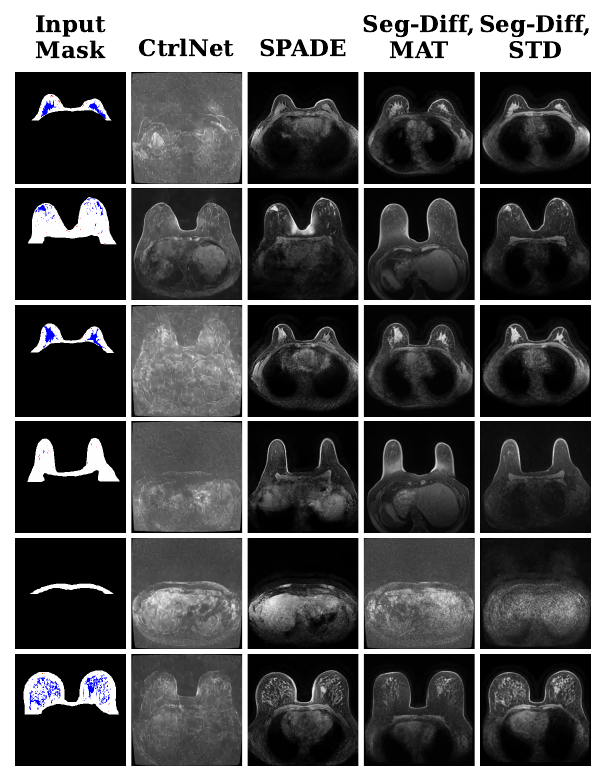}
    \includegraphics[width=0.44\linewidth]{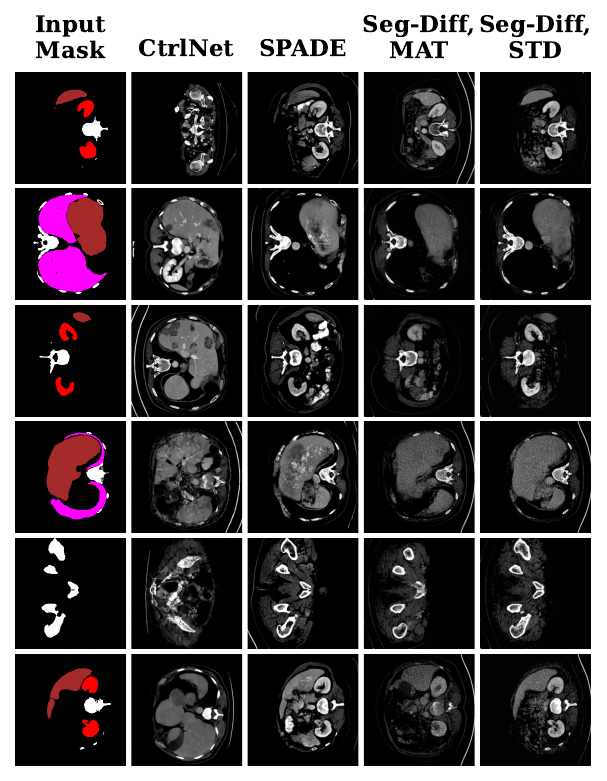}
   \caption{\textbf{Visual comparison of our model (SegGuidedDiff, or ``Seg-Diff'' for short) to existing segmentation-conditional image generation models.} For breast MRI, the breast, BV, and FGT segmentations are shown as white, red, and blue, respectively, while for CT, the liver, bladder, lungs, kidneys, and bone are in maroon, orange, pink, red, and white, respectively. ``MAT'' = max ablated training, ``STD'' = our standard method.
   }
    \label{fig:samples}
\end{figure}

\subsubsection{Evaluating Faithfulness of Generated Images to Input Masks.}

To measure how well our model follows an input mask for image generation, we use an auxiliary segmentation network trained on the real training set (a standard UNet; training details in Appendix A.2), to predict segmentations $m^{pred}_{gen}$ on images that were generated from the masks $m$ in the test set, and measure their overlap with (a) $m$ and (b) the model's predicted segmentations $m^{pred}_{real}$ for the input masks' original corresponding real images, similar to metrics used in \cite{park2019SPADE}. Our model's generated images have high overlap for both metrics ($>0.85$ Dice coeff., Table \ref{tab:faithful}), showing that our model closely followed the input masks when generating the images, and also outperformed the competing methods.

\subsubsection{Evaluating Generated Image Quality.}
We first attempted to use the common Fr\'echet Inception Distance (FID) \cite{fid} as a metric for quality/realism of generated image features compared to real data, via a CNN image encoder trained on the corresponding dataset. 
We observed that samples generated from both our segmentation-guided and standard unconditional diffusion models achieved potentially promising results, (\eg, breast MRI feature $\mathrm{FID}\simeq 0.5$), yet CNN feature-based metrics like FID fail to capture the global feature of anatomical realism that can differ in images generated by these models (\eg, fibroglandular tissue as shown in Fig. \ref{fig:eg_imgs}), so we caution using such metrics.

\begin{table}[htbp]
  \centering
  \caption{\textbf{Faithfulness of generated images to input masks.} $m$ denotes input masks, and $m^{pred}_{gen}$ and $m^{pred}_{real}$ denote the masks predicted for (a) the generated images and (b) the real images corresponding to the input masks, respectively, by an auxiliary segmentation model. Best-performing is shown in bold, and second best is underlined.}%
  \begin{tabular}{l||cc|cc}
  \toprule
  & \multicolumn{2}{c|}{\bfseries Breast MRI} & \multicolumn{2}{c}{\bfseries CT Organ} \\
  \toprule
  \multicolumn{1}{l||}{\bfseries Model} & \multicolumn{1}{c}{$\mathrm{Dice}(m^{pred}_{gen}, m)$} & \multicolumn{1}{c}{$\mathrm{Dice}(m^{pred}_{gen}, m^{pred}_{real})$} & \multicolumn{1}{|c}{$\mathrm{Dice}(m^{pred}_{gen}, m)$} & \multicolumn{1}{c}{$\mathrm{Dice}(m^{pred}_{gen}, m^{pred}_{real})$} \\ 
  \midrule
   ControlNet & 0.3636 & 0.3604 & 0.1132 & 0.1126  \\
   SPADE & \underline{0.8473} & \underline{0.8477} & \underline{0.8771} &  \underline{0.8603} \\
   \textbf{Ours} & \textbf{0.9027} & \textbf{0.8593} & \textbf{0.8980} & \textbf{0.8797}  \\
  \bottomrule
  \end{tabular}
  \label{tab:faithful}%
\end{table}

Instead, we propose to more precisely measure anatomical realism by determining how well the aforementioned auxiliary segmentation models for the objects of interest can be trained solely on these synthetic images to be able to generalize to real data, using the input masks as targets. We compare the performance of the segmentation models trained on (a) the real held-out training set (Sec. \ref{sec:data}) and (b) the set of images generated from all masks corresponding to these images. We split the real test set in half (by patient) into a validation set and a test set to use for these models. The results for this are in Table \ref{tab:qual}; we see that for both datasets, the segmentation network trained on our model's synthetic data barely performs worse (by only $\leq0.04$ Dice) than the network trained on real data, implying that our generated images are both highly realistic and faithful to the input masks (especially considering certain objects' segmentation difficulty). Our method is on par with SPADE for CT Organ, and slightly worse for breast MRI, while outperforming ControlNet in both cases.

\begin{table}[htbp]
  \centering
  \caption{\textbf{Quality of generated images.} Real test set performance (Dice coeff.) of a segmentation network for the objects of interest, trained on real data vs. synthetic data generated by different models.}
  \begin{tabular}{l|c|ccc}
  \toprule
  & \multirow{ 2}{*}{\textit{Real training set}} & \multicolumn{3}{c}{\textit{ Synthetic training set:}} \\
  & & ControlNet & SPADE & \textbf{Ours} \\
  \midrule
   Breast MRI & 0.8376 & 0.7570 & \textbf{0.8333} & \underline{0.7934} \\
   CT Organ & 0.9075 & 0.0000 & \underline{0.8932} & \textbf{0.8981} \\
  \bottomrule
  \end{tabular}
  \label{tab:qual}
\end{table}

ControlNet performed poorly for all metrics because it failed to follow input masks closely (if at all) due to its limitations and untested nature of adapting to medical images; more details are in Appendix A.1. We also see that our mask-ablated-trained model (Sec \ref{sec:mask_abla}) follows input masks less strictly than our standard model (Fig. \ref{fig:samples}), especially for the dataset with more object classes (CT Organ): for breast MRI the model obtained faithfulness metrics (Table \ref{tab:faithful}) of $\mathrm{Dice}(m^{pred}_{gen}, m) = 0.6589$ and $\mathrm{Dice}(m^{pred}_{gen}, m^{pred}_{real}) = 0.6684$, and quality metric (Table \ref{tab:qual}) of 0.7478, and 0.5952, 0.5963 and 0.7564 for CT Organ, respectively.
This is likely because the diversity of object class combinations seen in training scales exponentially with the number of classes in order to prepare the model for all possible combinations (Sec. \ref{sec:mask_abla}), so the model is ``spread thin''.
However, this relaxation of the mask constraint has its own benefits, detailed next.

\subsection{Advantages of Mask-Ablated-Training}
\label{sec:expmaskabla}

\subsubsection{Sampling from Ablated Masks.}
\begin{figure}[htbp]
\begin{minipage}[t]{0.45\textwidth}
    \centering
   \includegraphics[width=0.8\linewidth]{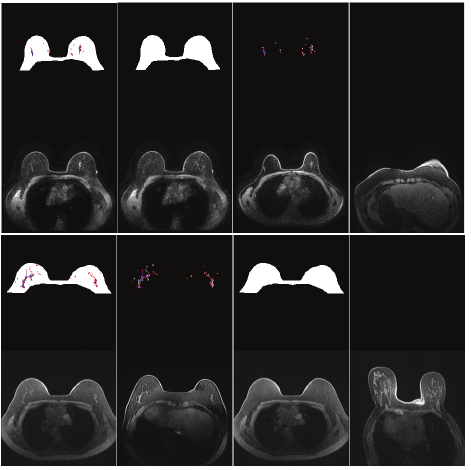}
   \caption{Generating images (even rows) from masks with classes removed (odd rows), shown for breast MRI.}
   \label{fig:samples_ablated}
\end{minipage}
\begin{minipage}[t]{0.5\textwidth}
    \centering
   \includegraphics[width=0.9\linewidth]{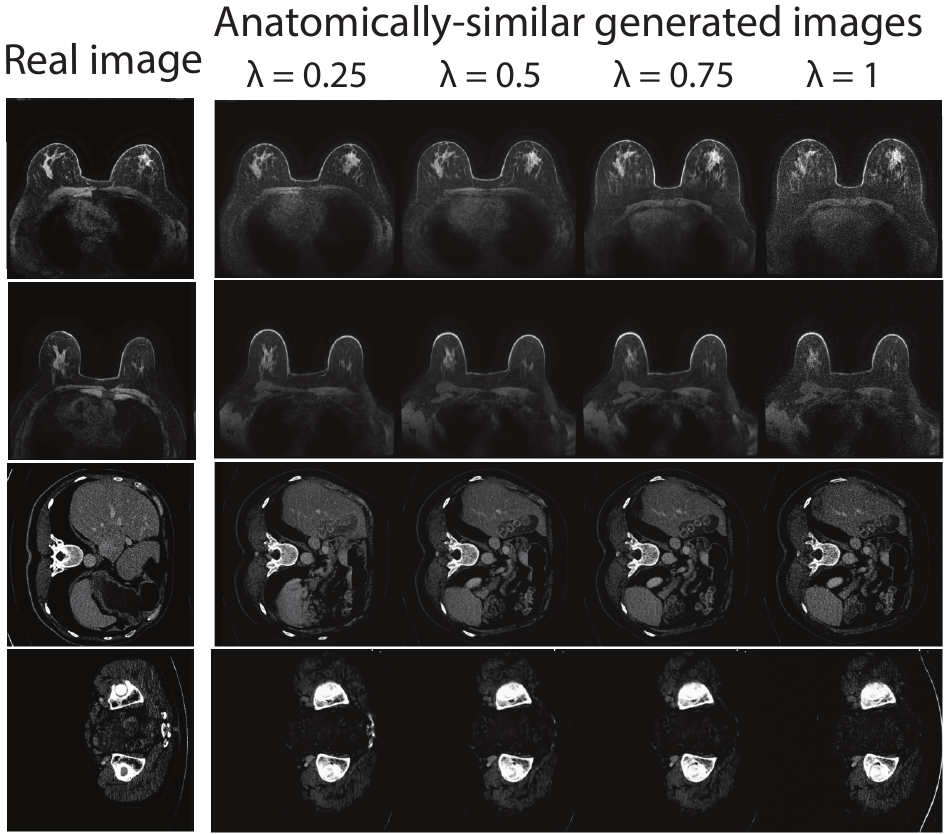}
   \caption{Using our model to generate images that are anatomically similar to real images.}
   \label{fig:sim}
\end{minipage}
\end{figure}

The direct benefit of mask ablated training (MAT) is its ability to generate images from masks with classes missing.
In Fig. \ref{fig:samples_ablated} we demonstrate the effect on generated images of ablating certain classes from an input mask for breast MRI (more examples, including for CT Organ, are provided in Appendix B).
For example, we see that constraining the BV+FGT in breast MRI, yet keeping the breast free, results in images that have the latter two classes pre-registered while the breast shape varies, and vice-versa.

\subsubsection{Adjustable Anatomical Similarity of Generated Images to Real Images.}

One application of our model is the adjustable generation of images that are anatomically similar to some real image, which is not immediately possible for existing state-of-the-art GAN-based methods like SPADE. Consider some real image $x_0$ with anatomical mask $m$. We can \textit{adjust} the anatomical similarity to $x_0$ of an image generated from $m$ with our model by interpolating between the synthetic image and the real image in the model's latent space, as follows. 

If the generation/de-noising process given $m$ (starting at timestep $t=T$) is halted at some intermediate step $t=\tilde{t}$ (we use $\tilde{t} = 240$), we obtain a latent representation $x'_{\tilde{t}}$ of the generating image. We can then convert the real image $x_0$ to the same latent space by applying the \textit{noising} process to $x_0$ from $t=0$ to $t=\tilde{t}$ to obtain $x_{\tilde{t}}$. Next, the features of the two images can be mixed via the \textit{interpolated} latent $x^\lambda_{\tilde{t}} := (1 - \lambda) x_{\tilde{t}} + \lambda x'_{\tilde{t}}$, where $\lambda \in (0,1]$ controls the similarity of the mixed features to those of the real image. From here, $x^\lambda_{\tilde{t}}$ can be denoised back to image space to obtain the interpolated image $x^\lambda_0$. An advantage of using our mask-ablated-trained model to generate $x^\lambda_0$ is that certain objects can be constrainted while others are free to vary, resulting in higher, yet adjustable, semantic diversity. We demonstrate this with various  $\lambda$ in Figure \ref{fig:sim}, with only FGT+BV constrained for breast MRI, and only bone constrained for CT Organ.

\section*{Conclusion}

Our segmentation-guided diffusion model enables superior anatomically-controllable medical image generation, which has many potential applications, including (1) the generation of anatomically rare cases to augment some imbalanced dataset, (2) the synthesis of anatomically-paired/pre-registered data, and (3) \textit{cross-modality} anatomy translation, where our model could be trained on images and masks from one sequence (\eg, T2 MRI), and then supplied with masks from another sequence (\eg, T1 MRI) to create new T2 images from the T1 masks.

However, some limitations of this study are that we did not consider full 3D generation, and we did not compare to ControlNet-like latent diffusion models \cite{stablediffusion} trained completely from scratch.
For future work, we are interested in further improving generated image quality, incorporating image-level class guidance \cite{ho2021classifier} either for pathological or domain-related variables \cite{konz2023reverse}, and extending our model to segmentation-guided image \textit{translation}.

\begin{credits}
\subsubsection{\ackname} Research reported in this publication was supported by the National Institute Of Biomedical Imaging And Bioengineering of the National Institutes of Health under Award Number R01EB031575. The content is solely the responsibility of the authors and does not necessarily represent the official views of the National Institutes of Health.

\subsubsection{\discintname} The authors have no competing interests.

\end{credits}

\bibliographystyle{splncs04}
\bibliography{main}

\clearpage
\appendix

\section{Additional Training and Architectural Details}
\label{app:implement}
\subsubsection{Our model (segmentation-guided diffusion).}
The denoising model (UNet)'s encoder is constructed from six standard ResNet down-sampling blocks, with the fifth block also having spatial self-attention, with $(128, 128, 256, 256, 512, 512)$ output channels, respectively. The decoder is simply the up-sampling reverse of the encoder. We use a standard forward process variance schedule that linearly increases from $\beta_1=10^{-4}$ to $\beta_T=0.02$ \cite{ho2020denoising}. For training, we use the AdamW optimizer \cite{adamw} and a cosine learning rate scheduler \cite{loshchilov2016sgdr} with an initial learning rate of $10^{-4}$, with $500$ linear warm-up steps. We train for 400 epochs with a batch size of 64 (about 26 hours), and we perform all training and evaluation on four 48 GB NVIDIA A6000 GPUs.
We use the Diffusers library as a backbone (\url{https://github.com/huggingface/diffusers}).

\subsection{Comparison models}
\label{app:allmodels}

\subsubsection{SPADE.}We train SPADE \cite{park2019SPADE} using the default settings, with a batch size of 128 for 50 epochs. We did not adopt the changes of the recent brain MRI SPADE model \cite{fernandez2022brainspade} because they are not applicable to our datasets/task, namely: (1) the contrast-based clustering is not applicable due to us using pre-contrast MRIs or CT, (2) we work with standard categorical segmentation maps, not partial volume/probabilistic segmentation maps, so changes using the latter are not applicable, and (3) we work with independent 2D slice images, rather than full 3D volumes, so the enforcement of style and content separation via using different slices from the same volume during training is not applicable.

\subsubsection{ControlNet.}
We adapted ControlNet \cite{controlnet} to each of our medical image datasets as was instructed at their official tutorial (\url{https://github.com/lllyasviel/ControlNet/blob/main/docs/train.md#sd_locked}) for use with datasets that are out-of-distribution (\eg, medical images) from their model's very large natural image pre-training set, using empty prompts for text inputs. We note that despite this tutorial, none of this was tested in the ControlNet paper, which may explain ControlNet's poor performance on our medical datasets.

This involved first finetuning the VAE for 200 epochs, then finetuning the Stable Diffusion (SD) model for 400 epochs using the respective breast MRI or CT organ training set images. 
We then finetuned the ControlNet with the images and their corresponding masks for segmentation guidance for 200 epochs. The pretrained (pre-finetuning) models are from the SD v1.5 checkpoints available on Hugging Face at \url{https://huggingface.co/runwayml/stable-diffusion-v1-5}. For all training, we set the batch size to 128, the initial learning rate to $10^{-4}$, and adopted cosine annealing learning rate schedulers rate with 500 steps of warm-up. 

\subsection{Auxiliary segmentation model}
\label{app:segmodel}
We used the MONAI UNet (\url{https://docs.monai.io/en/stable/networks.html}) with 1-channel input and (number of target object classes + 1)-channel output. The sequence of intermediate UNet channels was set to (16, 32, 64, 128, 256). We trained each model for 100 epochs with a batch size of 8 and selected the models with the lowest validation loss, with an initial learning rate of $10^{-3}$ and a cosine annealing scheduler.

\section{Additional Sampled Images}
\label{app:moresamples}

\begin{figure}[htbp]
    \centering
    \includegraphics[width=0.39\linewidth]{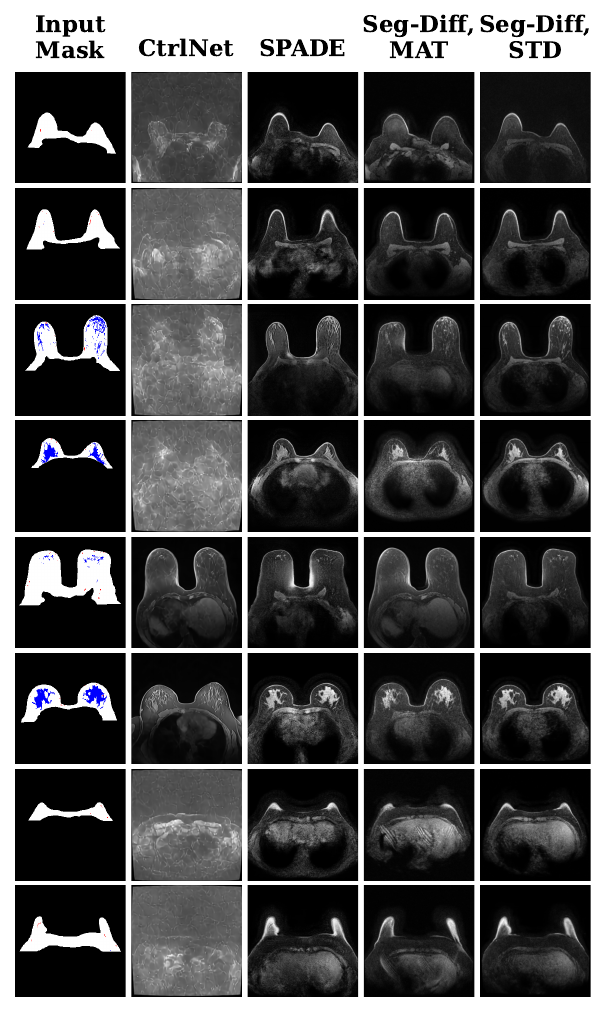}
    \includegraphics[width=0.39\linewidth]{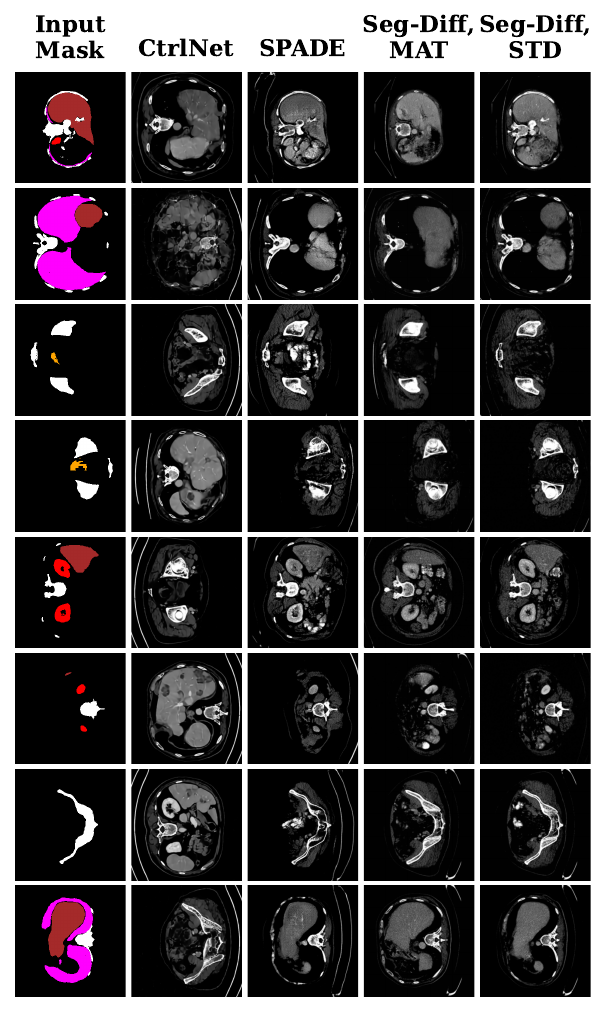}
   \caption{Additional samples from all segmentation-conditional models; breast MRI on the left, CT organ on the right. Please see Fig. \ref{fig:samples} caption for more details.}
   \label{fig:samples_many}
\end{figure}

\begin{figure}[htbp]
    
    \centering
    \includegraphics[width=0.4\linewidth]{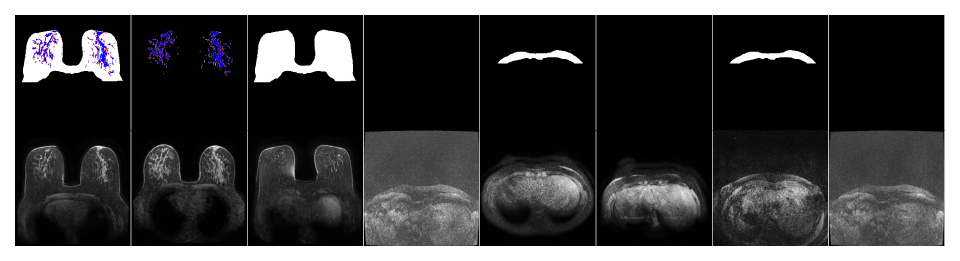}
    \includegraphics[width=0.4\linewidth]{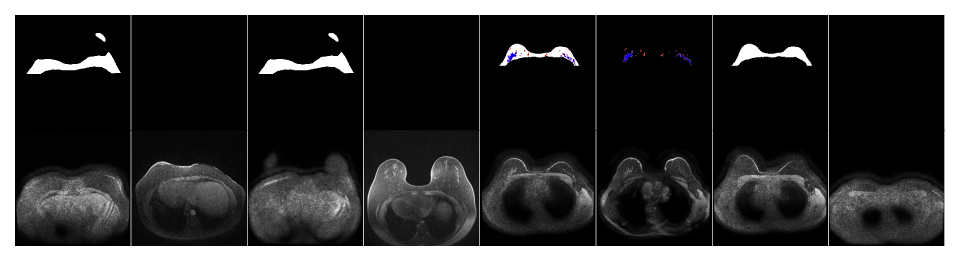}
    \includegraphics[width=0.4\linewidth]{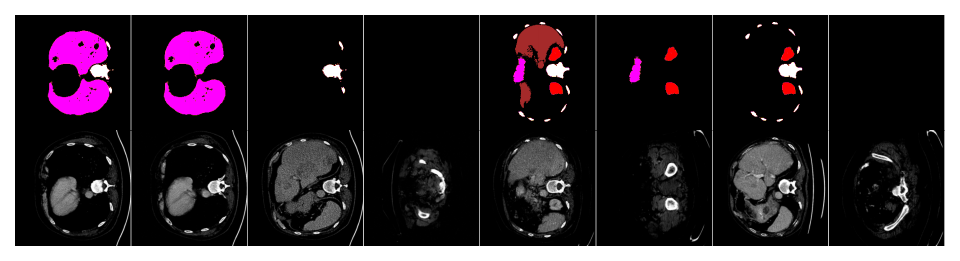}
    \includegraphics[width=0.4\linewidth]{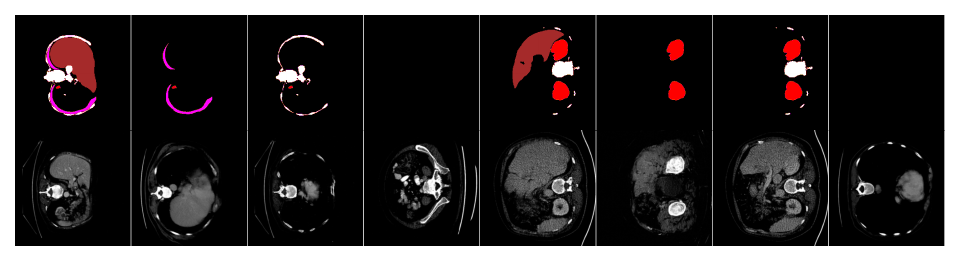}
   \caption{Additional samples from our mask-ablated-trained model with various classes removed from given input segmentations for breast MRI (top) and CT Organ (bottom).}
   \label{fig:samples_many_abla_breast}
\end{figure}

\end{document}